\documentclass[preprint,12pt]{elsarticle}

\usepackage{amssymb}
\usepackage{amsmath}
\usepackage{booktabs}
\usepackage{multirow}
\usepackage{tabularx}
\usepackage{array}
\usepackage{graphicx}
\usepackage{url}
\usepackage{orcidlink}

\begin{document}

\begin{frontmatter}



\title{MLFFM-SegDiff: A Multi-Level Feature Fusion Diffusion Model for Skin Lesion Segmentation} 


\author[inst1]{Jingjun Gu\corref{cor1}\orcidlink{0000-0001-5714-9189}}
\ead{gjj@zju.edu.cn}
\cortext[cor1]{Corresponding author}

\author[inst2]{Chaojie Shen}
\ead{qacket@126.com}

\author[inst2]{Yifeng Cao}
\ead{iyifeng@163.com}

\author[inst3]{Wei Zhang}
\ead{cstzhangwei@zju.edu.cn}

\author[inst4]{Yiliu Li}
\ead{lewislee@zju.edu.cn}

\author[inst1]{Aobo Fan}
\ead{fanab0528@ky.zstu.edu.cn}

\affiliation[inst1]{organization={Keyi College of Zhejiang Sci-Tech University},
            city={Shaoxing},
            country={China}}

\affiliation[inst2]{organization={Zhejiang Lanchen Digital Intelligence Technology Co., Ltd.},
            city={Hangzhou},
            country={China}}

\affiliation[inst3]{organization={School of Software Technology, Zhejiang University},
            city={Hangzhou},
            country={China}}

\affiliation[inst4]{organization={School of Medicine, Zhejiang University},
            city={Hangzhou},
            country={China}}

\begin{abstract}
Skin lesion segmentation is a fundamental task in computer-aided dermatological diagnosis, and its accuracy directly affects subsequent lesion analysis, feature extraction, and disease classification.
However, precise segmentation of dermoscopic images remains challenging because skin lesions often exhibit blurred boundaries, low contrast, large variations in shape and size, and interference from hair, shadows, and other background artifacts.
Recently, diffusion models have shown promising performance in medical image segmentation due to their progressive denoising and distribution modeling capabilities.
Nevertheless, existing diffusion-based skin lesion segmentation methods still have limitations in cross-level feature interaction and boundary detail recovery.

To address these issues, this paper proposes MLFFM-SegDiff, a multi-level feature fusion diffusion segmentation model for skin lesion segmentation.
Built upon a diffusion-based segmentation framework, the proposed method improves the denoising network by introducing a dual-path U-Net encoder, a Multi-Level Feature Fusion Module (MLFFM), and a configurable boundary-sensitive loss term.
The dual-path encoder strengthens the interaction between noisy mask features and dermoscopic image features, while MLFFM enhances skip-connection features through attention mechanisms, scale alignment, and adaptive cross-level fusion.
In this way, the decoder can jointly exploit shallow boundary details and deep semantic representations, thereby improving the reconstruction quality of lesion masks.

Experiments are conducted on three public skin lesion datasets, namely ISIC2018, PH2, and HAM10000.
The proposed method is compared with several representative segmentation models, including DermoSegDiff, Generic U-Net, SwinUNETR, and U-Net.
Experimental results show that MLFFM-SegDiff achieves the best average performance across the three datasets in terms of Accuracy, F1-score, Jaccard index, Recall, and Dice coefficient.
In particular, the average Jaccard index and Dice coefficient reach \textbf{0.8546} and \textbf{0.9207}, respectively.
These results demonstrate that the proposed multi-level feature fusion mechanism effectively improves lesion region overlap and recall performance, providing a feasible and effective approach for automatic skin lesion segmentation.
The source code will be made publicly available in the GitHub repository at \url{https://github.com/Qacket/MLFFM-SegDiff.git} following the publication of this article.
\end{abstract}


\end{frontmatter}

\section{Introduction}

Skin diseases are common in clinical practice, among which malignant lesions such as melanoma can pose serious health risks if they are not detected and treated in time \cite{brinker2019,esteva2017}.
With the development of digital dermoscopy and medical artificial intelligence, image-based computer-aided diagnosis has become an important tool for dermatological screening and clinical decision support.
In such systems, automatic skin lesion segmentation is a fundamental step for subsequent feature extraction, area measurement, boundary analysis, and lesion classification.
Accurate segmentation can reduce interference from surrounding normal skin and background regions, while providing reliable support for quantitative analysis of lesion morphology, color, and texture.
Therefore, developing robust and accurate skin lesion segmentation models is of great research significance and practical value.

Although deep learning has substantially improved medical image segmentation, accurate skin lesion segmentation remains challenging.
First, lesion regions often have blurred or ambiguous boundaries, especially when color transitions are smooth or lesion contours are irregular \cite{celebi2019}.
Second, variations in patients, imaging devices, and illumination conditions can cause significant differences in color, brightness, and texture distributions, making model generalization more difficult.
In addition, artifacts such as hair, shadows, reflections, and low-contrast regions may further interfere with lesion localization.
Since skin lesions usually exhibit large variations in shape and scale, features from a single level or scale are often insufficient to simultaneously capture fine boundary details and high-level semantic structures.

Existing skin lesion segmentation methods are mainly based on convolutional neural networks, Transformer architectures, or diffusion models.
Encoder-decoder networks represented by U-Net have been widely used in medical image segmentation because skip connections can combine shallow spatial details with deep semantic information \cite{ronneberger2015unet}.
However, conventional skip connections usually transfer features only between corresponding encoder and decoder levels, resulting in limited cross-level interaction and multi-scale feature fusion.
When dealing with skin lesions with fuzzy boundaries, irregular shapes, and large scale variations, such a feature transfer strategy may lead to insufficient boundary recovery or inadequate semantic utilization.
Transformer-based methods enhance global dependency modeling through self-attention mechanisms \cite{dosovitskiy2021,hatamizadeh2022swinunetr}, but they often require large-scale training data and high computational resources, and may suffer from limited stability and generalization on small medical datasets.
Recently, diffusion models have been introduced into image segmentation tasks \cite{ho2020ddpm, song2021ddim}.
By progressively recovering target masks through forward noising and reverse denoising processes, diffusion models provide a promising framework for modeling segmentation uncertainty and complex lesion boundaries.
Methods such as DermoSegDiff have demonstrated the effectiveness of diffusion models for skin lesion segmentation.
Nevertheless, existing diffusion-based methods still have room for improvement.

To address these limitations, this paper proposes MLFFM-SegDiff, a multi-level feature fusion diffusion model for skin lesion segmentation.
Built upon the DermoSegDiff framework, the proposed method improves the denoising network from three aspects: feature extraction, cross-level feature fusion, and configurable boundary-sensitive optimization.
First, a dual-path U-Net encoder is introduced to enhance the interaction between original image features and noisy mask features during feature extraction.
Second, a Multi-Level Feature Fusion Module (MLFFM) is designed between the encoder and decoder to jointly model skip features from different encoder levels.
Finally, a boundary-sensitive loss term is included as a configurable training component to assign higher weights to boundary regions, encouraging the model to pay more attention to ambiguous lesion contours when this term is enabled.

The proposed method is evaluated on three public skin lesion datasets, namely ISIC2018, PH2, and HAM10000 \cite{codella2019isic, tschandl2018ham10000, mendonca2013ph2}.
It is compared with representative segmentation methods, including DermoSegDiff, Generic U-Net, SwinUNETR, and U-Net.
Experimental results show that MLFFM-SegDiff achieves superior average performance across the three datasets in terms of Accuracy, F1-score, Jaccard index, Recall, and Dice coefficient.
These results demonstrate the effectiveness of the proposed method in improving lesion localization, region coverage, and overlap quality.

The main contributions of this paper are summarized as follows:

\begin{enumerate}
  \item A multi-level feature fusion diffusion model, named MLFFM-SegDiff, is proposed for skin lesion segmentation, improving feature extraction, cross-level interaction, and boundary-aware optimization.

  \item A dual-path U-Net encoder, a Multi-Level Feature Fusion Module (MLFFM), and a configurable boundary-sensitive loss are designed to enhance semantic and boundary representation.

  \item Extensive experiments on ISIC2018, PH2, and HAM10000 demonstrate that the proposed method outperforms representative baselines on multiple evaluation metrics.
\end{enumerate}

In summary, this paper improves automatic skin lesion segmentation by introducing a dual-path feature extraction structure, a multi-level feature fusion mechanism, and a boundary-sensitive optimization strategy into a diffusion-based segmentation framework.
The proposed method provides a useful reference for structural optimization of diffusion models in medical image segmentation and offers a promising solution for robust skin lesion segmentation.

\section{Related Work}

\subsection{Skin Lesion and Medical Image Segmentation}

Medical image segmentation has long been an important topic in computer-aided diagnosis.
Fully convolutional networks established end-to-end dense prediction as a fundamental paradigm for semantic segmentation \cite{long2015fcn}.
U-Net introduced an encoder-decoder architecture with skip connections, allowing contextual semantics and spatial details to be effectively combined, and has become a classical baseline for biomedical image segmentation \cite{ronneberger2015unet}.
Subsequent studies further improved U-shaped segmentation networks by introducing attention mechanisms to highlight foreground regions and suppress irrelevant background responses.
For example, Attention U-Net incorporated attention gates into skip connections to improve target localization in medical images \cite{oktay2018attunet}.
Other U-shaped variants, such as UNet++ and nnU-Net, further explored nested skip pathways and self-configuring training pipelines to improve robustness across biomedical segmentation tasks \cite{zhou2018unetpp,isensee2021nnunet}.
Nevertheless, skin lesion segmentation remains challenging because lesion boundaries are often ambiguous, lesion scales vary substantially, and background artifacts such as hair or shadows may interfere with prediction.
In such cases, conventional same-level skip connections may be insufficient for integrating multi-level semantic and boundary information.

Compared with general semantic segmentation, skin lesion segmentation relies more heavily on boundary details and region overlap quality.
Lesion regions often exhibit irregular shapes, smooth color transitions, and texture similarity with surrounding normal skin.
Therefore, relying only on local convolutional features may lead to under-segmentation or over-segmentation along lesion boundaries.
This motivates segmentation models to better combine local spatial details, high-level semantic representations, and boundary-sensitive information.

\subsection{Transformer-Based Medical Image Segmentation}

Transformers enhance long-range dependency modeling through self-attention and have been increasingly adopted in medical image segmentation.
TransUNet combines CNN features with a Transformer encoder to extract global contextual information for medical image segmentation \cite{chen2021transunet}.
Swin UNETR employs a hierarchical Swin Transformer encoder to model multi-scale representations and has demonstrated strong performance in volumetric medical image segmentation \cite{hatamizadeh2022swinunetr}.
These methods compensate for the limited global modeling capacity of CNNs, but they usually introduce higher computational cost and may require careful training design when annotated medical datasets are limited.

Although Transformer-based models strengthen global semantic representation, skin lesion segmentation still requires accurate recovery of local boundaries.
For low-contrast lesions or lesions with blurred contours, enhancing global dependencies alone may not fully resolve the loss of boundary details.
Therefore, how to preserve local spatial details while introducing richer cross-level semantic information remains a key problem in skin lesion segmentation model design.

\subsection{Diffusion Models for Segmentation}

Diffusion models first achieved remarkable progress in generative modeling.
Denoising Diffusion Probabilistic Models (DDPMs) learn data distributions through a forward noising process and a reverse denoising process, providing a new modeling paradigm for image generation and downstream vision tasks \cite{ho2020ddpm}.
To reduce the high inference cost of iterative denoising, subsequent diffusion studies investigated faster sampling strategies and improved reverse-process parameterization \cite{song2021ddim,nichol2021improved}.
Later studies extended diffusion models to segmentation.
SegDiff formulated mask prediction as an iterative denoising process and fused multiple stochastic samples to obtain the final segmentation map \cite{segdiff2021}.
MedSegDiff applied diffusion probabilistic models to medical image segmentation and introduced dynamic conditional encoding and a Feature Frequency Parser to enhance regional attention during step-wise denoising \cite{medsegdiff2022}.

In skin lesion segmentation, DermoSegDiff incorporated boundary information into a diffusion-based segmentation framework and demonstrated the potential of diffusion models for lesion boundary modeling \cite{dermosegdiff2023}.
These studies indicate that boundary priors, scale interactions, and stable denoising strategies are important for diffusion-based skin lesion segmentation.
However, existing diffusion-based segmentation methods still face several limitations.
First, the interaction between original image features and noisy mask features may be insufficient when simple encoder structures are used.
Second, the multi-level features between the encoder and decoder are not always fully exploited.
Third, boundary details and global semantic information may not be jointly optimized in a sufficiently effective manner.
These limitations leave room for further improving the denoising network structure and optimization strategy of diffusion-based skin lesion segmentation models.

\subsection{Multi-Level Feature Fusion and Boundary-Aware Learning}

Multi-level feature fusion is an important strategy in medical image segmentation.
Shallow features usually contain edge, texture, and spatial location information, whereas deep features provide stronger semantic representation.
For skin lesion images, both lesion boundaries and global lesion structures are important, and therefore models need to exploit information from different levels simultaneously.
Feature pyramid and lateral-connection designs have shown the value of multi-scale feature aggregation in dense prediction tasks \cite{lin2017fpn}.
If skip connections transfer only single-level features, the decoder may fail to obtain sufficiently complete cross-scale context.
Attention-based refinement is also useful for emphasizing informative channels and spatial regions, and channel-spatial attention modules have been widely used as lightweight feature enhancement blocks \cite{woo2018cbam}.

Boundary-aware learning strengthens the model's focus on lesion contours from another perspective.
Existing diffusion-based skin lesion segmentation studies suggest that incorporating boundary information into the denoising process can improve boundary recovery.
Boundary-based objectives further show that contour-oriented constraints can complement region-based segmentation losses, especially when foreground and background regions are highly imbalanced \cite{kervadec2021boundary}.
In this paper, MLFFM-SegDiff follows this research direction and further improves the diffusion segmentation framework through dual-path feature extraction, multi-level skip feature fusion, and configurable boundary-sensitive optimization.
The dual-path U-Net encoder enhances the interaction between original image features and noisy mask features.
The Multi-Level Feature Fusion Module (MLFFM) applies attention enhancement, scale alignment, and cross-level fusion to features from different encoder levels.
In addition, the boundary-sensitive loss function assigns higher weights to boundary regions when enabled, encouraging the model to focus on ambiguous lesion contours during training.
Through these designs, the proposed method aims to improve the joint representation of boundary details and semantic structures, thereby enhancing lesion overlap quality and recall performance.

\section{Method}

\subsection{Problem Formulation}

Given a dermoscopic image $I \in \mathbb{R}^{H \times W \times 3}$, the goal of skin lesion segmentation is to predict its corresponding binary lesion mask $x_0 \in \{0,1\}^{H \times W}$.
Following diffusion-based segmentation frameworks, this paper formulates skin lesion segmentation as a conditional denoising process, where the dermoscopic image is used as the guidance condition and the lesion mask is recovered through iterative reverse diffusion.

In the forward diffusion process, Gaussian noise is gradually added to the ground-truth mask $x_0$ according to a predefined noise schedule.
At timestep $t$, the noisy mask $x_t$ is generated as:
\begin{equation}
x_t = \sqrt{\bar{\alpha}_t}x_0 + \sqrt{1-\bar{\alpha}_t}\epsilon,
\quad \epsilon \sim \mathcal{N}(0,\mathbf{I}),
\end{equation}
where $\bar{\alpha}_t$ is determined by the diffusion schedule.
The denoising network $f_{\theta}$ takes the noisy mask $x_t$, the guidance image $I$, and the timestep $t$ as inputs, and predicts the injected noise:
\begin{equation}
\hat{\epsilon}_{\theta} = f_{\theta}(x_t, I, t).
\end{equation}
During inference, the predicted noise is used to progressively remove noise from the mask distribution, and the final lesion mask is obtained after the reverse denoising process.

\subsection{Overall Architecture}

\begin{figure*}[htb]
  \centering
  \includegraphics[width=\linewidth]{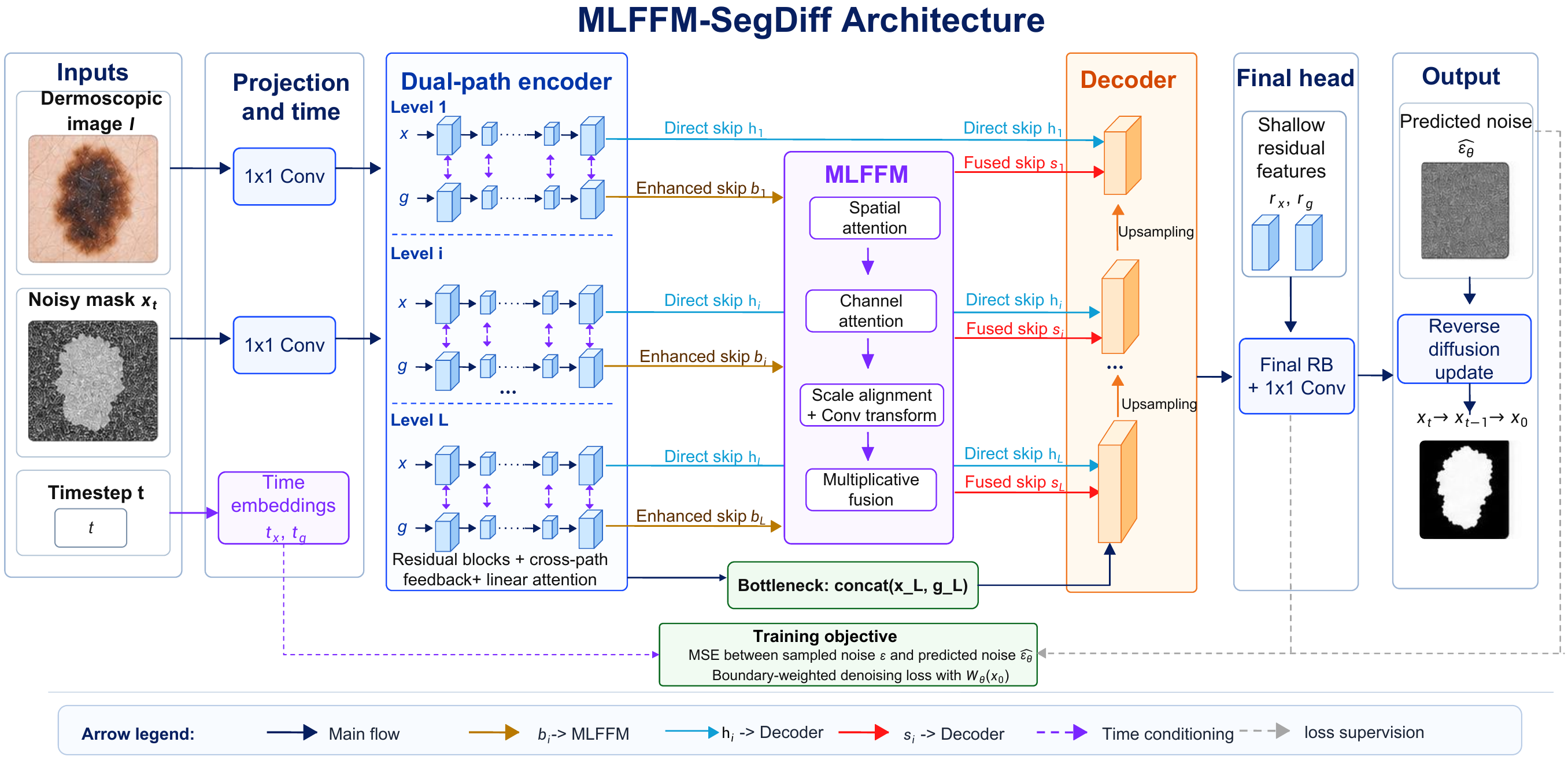}
  \caption{
  Overall architecture of MLFFM-SegDiff. The noisy mask and dermoscopic image are encoded by a dual-path encoder, where enhanced skip features are fused by MLFFM and direct skip features are preserved for the decoder. The bottom legend summarizes the arrow semantics.
  }
  \label{fig:framework}
\end{figure*}

The overall architecture of MLFFM-SegDiff is shown in Fig.~\ref{fig:framework}.
The model follows a conditional diffusion segmentation framework in which the noisy mask $x_t$, the dermoscopic image $I$, and the timestep $t$ are jointly used to predict the injected noise.
The network contains four main components: input projection layers, a dual-path encoder, the proposed Multi-Level Feature Fusion Module (MLFFM), and a decoder for mask reconstruction.

At the input stage, the noisy mask and the image guidance branch are mapped into feature spaces by separate $1 \times 1$ convolutions.
The dual-path encoder then extracts mask-related and image-guided representations in parallel and establishes conditional interaction between the two branches.
During encoding, two types of skip features are preserved.
The direct skip features $h_i$ retain local encoder information, while the guidance-enhanced skip features $b_i$ are sent to MLFFM for attention enhancement, scale alignment, and cross-level fusion.
The fused skip features $s_i$ are then combined with the direct skip features in the decoder, allowing the reconstruction process to exploit both boundary details from shallow layers and semantic structures from deeper layers.
Finally, the decoder output is fused with the initial shallow features and mapped to the predicted noise $\hat{\epsilon}_{\theta}$ for reverse diffusion.

\subsection{Dual-Path Encoder}

In the original diffusion-based segmentation framework, the noisy mask and the guidance image are usually fused using relatively simple feature interaction strategies.
However, for skin lesion images, the lesion boundaries are often blurred and the appearance difference between lesion and normal skin can be subtle.
Simple feature fusion may therefore be insufficient to capture both lesion semantics and boundary details.
To address this issue, MLFFM-SegDiff adopts a dual-path encoder to jointly model the noisy mask branch and the image guidance branch.

Specifically, the noisy mask $x_t$ and the guidance image $I$ are first mapped by separate $1 \times 1$ convolutions.
The obtained shallow features are denoted as $x$ and $g$, respectively.
Meanwhile, the timestep $t$ is encoded by sinusoidal position embeddings and multilayer perceptrons to generate time embeddings for the two branches.
The shallow features are also preserved as $r_x$ and $r_g$ for the final prediction stage.

At the $i$-th encoder module, the mask feature and image feature are first processed by residual blocks:
\begin{equation}
\hat{x}_i = \mathrm{RB}^{x}_{i,1}(x_{i-1}, t_x),
\quad
\hat{g}_i = \mathrm{RB}^{g}_{i,1}(g_{i-1}, t_g),
\end{equation}
where $t_x$ and $t_g$ denote the time embeddings for the mask branch and guidance branch, respectively.
The two features are then concatenated to establish conditional interaction:
\begin{equation}
h_i = [\hat{x}_i, \hat{g}_i],
\end{equation}
where $[\cdot,\cdot]$ denotes channel-wise concatenation.
The concatenated feature is further processed by a residual block in the guidance branch:
\begin{equation}
\bar{g}_i = \mathrm{RB}^{g}_{i,2}(h_i,t_g).
\end{equation}
Then, $\bar{g}_i$ is projected by a $1 \times 1$ convolution and used to modulate the mask feature:
\begin{equation}
\bar{x}_i =
\mathrm{RB}^{x}_{i,2}
\left(
\eta_i(\bar{g}_i) \odot \hat{x}_i, t_x
\right),
\end{equation}
where $\eta_i(\cdot)$ denotes the $1 \times 1$ feedback convolution and $\odot$ denotes element-wise multiplication.
Finally, linear attention is applied to refine the two branches:
\begin{equation}
x_i = \mathrm{LAtt}(\bar{x}_i),
\quad
b_i = [x_i,\bar{g}_i],
\quad
g_i = \mathrm{LAtt}(b_i).
\end{equation}
The encoder module outputs the updated features $(x_i,g_i)$ and two skip features $(h_i,b_i)$.
Here, $h_i$ is directly preserved for decoder reconstruction, while $b_i$ is further enhanced by the proposed multi-level feature fusion module.

\subsection{Multi-Level Feature Fusion Module}

The multi-level feature fusion module is designed to improve the skip connections between the encoder and decoder.
In conventional U-Net-like networks, skip connections usually transfer only corresponding-level features.
This strategy may limit cross-level semantic interaction, especially for skin lesion images with irregular shapes, scale variations, and fuzzy boundaries.
In MLFFM-SegDiff, the skip features $\{b_i\}_{i=1}^{L}$ generated by the dual-path encoder are first enhanced and then fused across multiple levels.

For each skip feature $b_i$, a SkipBlock is used for attention enhancement.
The SkipBlock consists of spatial attention, channel attention, and a convolutional transformation.
The attention enhancement process can be written as:
\begin{equation}
b_i^{SA} = \mathcal{A}_s(b_i) \odot b_i,
\end{equation}
\begin{equation}
b_i^{SA\cdot CA} = \mathcal{A}_c(b_i^{SA}) \odot b_i^{SA},
\end{equation}
where $\mathcal{A}_s(\cdot)$ and $\mathcal{A}_c(\cdot)$ denote spatial attention and channel attention, respectively.
Through this process, informative spatial regions and channels are emphasized before cross-level fusion.

After attention enhancement, a scale-aligned multiplicative fusion operation is applied.
For the $i$-th decoder level, $b_i^{SA\cdot CA}$ is used as the anchor feature.
All enhanced skip features are aligned to the spatial size of the anchor feature.
Features with larger resolution are downsampled by adaptive average pooling, while features with smaller resolution are upsampled by bilinear interpolation.
The aligned features are then transformed and fused by element-wise multiplication:
\begin{equation}
s_i =
\operatorname{ReLU}
\left(
\prod_{j=1}^{L}
C_{ij}
\left(
\operatorname{Align}
\left(
b_j^{SA\cdot CA}, b_i^{SA\cdot CA}
\right)
\right)
\right),
\end{equation}
where $s_i$ denotes the fused skip feature for the $i$-th decoder stage, $\operatorname{Align}(\cdot,\cdot)$ denotes the scale-alignment operation, and $C_{ij}(\cdot)$ denotes the convolutional transformation that maps the aligned feature to the required channel dimension.
By using this module, each decoder stage can obtain cross-level information rather than relying only on a single corresponding skip feature.
Shallow features provide boundary and texture details, while deeper features provide global semantic information, which helps improve lesion mask reconstruction.

\subsection{Bottleneck and Decoder}

After the final encoder stage, the mask branch and guidance branch features are concatenated and passed into the bottleneck module.
The bottleneck module consists of residual blocks and a hybrid attention operation that combines self-attention and linear attention:
\begin{equation}
z = \mathcal{B}([x_L,g_L],t_x),
\end{equation}
where $z$ denotes the bottleneck feature and $\mathcal{B}(\cdot)$ represents the bottleneck module.

The decoder progressively upsamples the bottleneck feature and reconstructs the lesion mask representation.
At each decoder stage, the decoder receives three types of information: the current decoder feature, the direct skip feature $h_i$, and the fused skip feature $s_i$ generated by MLFFM.
Consistent with the implementation, the decoder first concatenates the current feature with $s_i$ and processes it using a residual block.
It then concatenates the result with $h_i$ and applies another residual block followed by linear attention:
\begin{equation}
d_i = \mathcal{D}_i(d_{i+1}, s_i, h_i, t_x),
\end{equation}
where $d_{i+1}$ denotes the feature from the deeper decoder stage and $\mathcal{D}_i(\cdot)$ denotes the decoder module.

After the final decoder stage, the decoded feature is concatenated with the shallow mask feature $r_x$ and the shallow image feature $r_g$ preserved from the input projection stage.
A final residual block and a $1 \times 1$ convolution are used to output the predicted noise:
\begin{equation}
\hat{\epsilon}_{\theta}
=
\operatorname{Conv}_{1\times1}
\left(
\mathcal{R}([d_1,r_x,r_g],t_x)
\right),
\end{equation}
where $\mathcal{R}(\cdot)$ denotes the final residual block.
The predicted noise $\hat{\epsilon}_{\theta}$ is then used in the reverse diffusion process to recover the final segmentation mask.

\subsection{Training Objective}

The basic training objective of MLFFM-SegDiff follows the standard noise prediction loss in diffusion models.
Given the sampled noise $\epsilon$ and the predicted noise $\hat{\epsilon}_{\theta}$, the denoising loss is defined as:
\begin{equation}
\mathcal{L}_{\mathrm{denoise}}
=
\mathbb{E}_{x_0,I,t,\epsilon}
\left[
\left\|
\epsilon - \hat{\epsilon}_{\theta}(x_t,I,t)
\right\|_2^2
\right].
\end{equation}

To strengthen the model's attention to lesion boundaries, the implementation also supports a boundary-sensitive weighting strategy.
A boundary attention map $W_{\Theta}(x_0)$ is generated from the ground-truth mask according to the distance between each pixel and the lesion contour.
Pixels closer to the boundary are assigned higher weights, while pixels farther from the boundary receive lower weights.
The time-dependent boundary weighting coefficient is defined as:
\begin{equation}
\gamma(t) = \gamma_0 e^{-\mu t},
\end{equation}
where $\gamma_0$ denotes the initial boundary weight and $\mu$ controls the decay rate.

The final training objective is written as:
\begin{equation}
\mathcal{L}
=
\mathbb{E}_{x_0,I,t,\epsilon}
\left[
\left(
1 + \lambda_b \gamma(t) W_{\Theta}(x_0)
\right)
\odot
\left(
\epsilon -
\hat{\epsilon}_{\theta}(x_t,I,t)
\right)^2
\right],
\end{equation}
where $\lambda_b$ controls the strength of the boundary-sensitive term.
When the boundary-sensitive term is not enabled in the configuration, $\lambda_b$ is set to zero.
This training objective encourages the model to maintain stable denoising performance while improving its sensitivity to ambiguous lesion boundaries.

\section{Experiments}

\subsection{Datasets}

The proposed method is evaluated on three public skin lesion datasets, including ISIC2018, PH2, and HAM10000.
ISIC2018 is a skin lesion analysis challenge dataset provided by the International Skin Imaging Collaboration, containing dermoscopic images and corresponding lesion annotations \cite{codella2019isic}.
PH2 is a dermoscopic image database with expert-annotated lesion masks and is widely used for evaluating skin lesion segmentation algorithms \cite{mendonca2013ph2}.
HAM10000 is a large-scale multi-source dermoscopic image collection of pigmented skin lesions \cite{tschandl2018ham10000}.

All datasets are formulated as binary segmentation tasks, where lesion regions are regarded as foreground and non-lesion regions are regarded as background.
In our experiments, all input images and masks are resized to $128 \times 128$ before being fed into the network.
The same preprocessing and evaluation protocol is applied to all comparison methods to ensure a fair comparison.
To obtain robust performance estimates, a five-fold cross-validation strategy is adopted, where four folds are used for training and the remaining fold is used for testing in each iteration.
The final results are reported as the average performance over all five folds.

\subsection{Implementation Details}

All experiments are implemented using PyTorch.
MLFFM-SegDiff adopts a dual-branch architecture, where the noisy mask branch takes the segmentation mask as input and the image guidance branch uses the corresponding RGB image.
The diffusion process follows a linear noise schedule, and the model is optimized with Adam together with a ReduceLROnPlateau learning-rate scheduler.

Dataset-specific training epochs and batch sizes are selected according to the scale of each dataset.
During inference, multiple stochastic sampling results are ensembled to reduce the randomness introduced by the diffusion process.

Both pixel-level and spatial data augmentations are applied, including intensity perturbation, color adjustment, geometric transformation, distortion, and random dropout.
The boundary-weighted loss is treated as a configurable training component and is enabled for datasets where boundary modeling is beneficial.

The comparison methods include DermoSegDiff, Generic U-Net, SwinUNETR, and U-Net.
All models are evaluated using the same metrics, and the quantitative results are reported as mean $\pm$ standard deviation.

\subsection{Evaluation Metrics}

Seven commonly used metrics are adopted to evaluate segmentation performance, including Accuracy, F1-score, Jaccard index, Precision, Recall, Specificity, and Dice coefficient.
Let TP, TN, FP, and FN denote true positives, true negatives, false positives, and false negatives, respectively.
The metrics are defined as:

\begin{align}
    \mathrm{Accuracy} &= \frac{\mathrm{TP}+\mathrm{TN}}{\mathrm{TP}+\mathrm{TN}+\mathrm{FP}+\mathrm{FN}},\\
    \mathrm{Precision} &= \frac{\mathrm{TP}}{\mathrm{TP}+\mathrm{FP}},\\
    \mathrm{Recall} &= \frac{\mathrm{TP}}{\mathrm{TP}+\mathrm{FN}},\\
    \mathrm{Specificity} &= \frac{\mathrm{TN}}{\mathrm{TN}+\mathrm{FP}},\\
    \mathrm{F1} &= \frac{2\cdot\mathrm{Precision}\cdot\mathrm{Recall}}{\mathrm{Precision}+\mathrm{Recall}},\\
    \mathrm{Jaccard} &= \frac{\mathrm{TP}}{\mathrm{TP}+\mathrm{FP}+\mathrm{FN}},\\
    \mathrm{Dice} &= \frac{2\mathrm{TP}}{2\mathrm{TP}+\mathrm{FP}+\mathrm{FN}}.
\end{align}
For binary segmentation, F1-score and Dice coefficient are closely related and may become equivalent under the same averaging strategy.
Both metrics are reported to remain consistent with common skin lesion segmentation evaluation protocols.

\subsection{Quantitative Results}

Table~\ref{tab:main-results} presents the quantitative comparison results on ISIC2018, PH2, and HAM10000.
On ISIC2018, MLFFM-SegDiff achieves the best Accuracy, F1-score, Jaccard index, Recall, and Dice coefficient, while SwinUNETR and U-Net obtain slightly higher Precision and Specificity.
On PH2 and HAM10000, the proposed method achieves the best performance across all seven evaluation metrics.

These results show that MLFFM-SegDiff improves lesion overlap and region coverage, especially in terms of Jaccard, Recall, and Dice.
This suggests that the dual-path feature interaction and multi-level feature fusion are effective for lesion mask reconstruction.

\begin{table*}[t]
\centering
\caption{Quantitative comparison on ISIC2018, PH2, and HAM10000. Results are reported as mean $\pm$ standard deviation.}
\label{tab:main-results}
\scriptsize
\resizebox{\textwidth}{!}{%
\begin{tabular}{llccccccc}
\toprule
Dataset & Model & Acc. & F1 & Jaccard & Precision & Recall & Specificity & Dice \\
\midrule
ISIC2018 & DermoSegDiff & $0.8895{\pm}0.1089$ & $0.7955{\pm}0.1213$ & $0.6758{\pm}0.1534$ & $0.8350{\pm}0.2175$ & $0.8240{\pm}0.1153$ & $0.9064{\pm}0.1615$ & $0.7955{\pm}0.1213$ \\
ISIC2018 & Generic U-Net & $0.8495{\pm}0.0105$ & $0.6662{\pm}0.0144$ & $0.4997{\pm}0.0162$ & $0.6382{\pm}0.0437$ & $0.7019{\pm}0.0373$ & $0.8900{\pm}0.0233$ & $0.6662{\pm}0.0145$ \\
ISIC2018 & SwinUNETR & $0.9486{\pm}0.0030$ & $0.8782{\pm}0.0078$ & $0.7830{\pm}0.0124$ & $\mathbf{0.8910{\pm}0.0069}$ & $0.8661{\pm}0.0174$ & $\mathbf{0.9710{\pm}0.0034}$ & $0.8782{\pm}0.0078$ \\
ISIC2018 & U-Net & $0.9483{\pm}0.0025$ & $0.8784{\pm}0.0051$ & $0.7832{\pm}0.0081$ & $0.8843{\pm}0.0037$ & $0.8726{\pm}0.0084$ & $0.9689{\pm}0.0015$ & $\mathbf{0.8784{\pm}0.0051}$ \\
ISIC2018 & MLFFM-SegDiff & $\mathbf{0.9504{\pm}0.0001}$ & $\mathbf{0.8877{\pm}0.0202}$ & $\mathbf{0.7985{\pm}0.0008}$ & $0.8588{\pm}0.0013$ & $\mathbf{0.9199{\pm}0.0003}$ & $0.9588{\pm}0.0001$ & $\mathbf{0.8877{\pm}0.0202}$ \\
\midrule
PH2 & DermoSegDiff & $0.8850{\pm}0.0237$ & $0.8235{\pm}0.0214$ & $0.7005{\pm}0.0312$ & $0.8319{\pm}0.0758$ & $0.8252{\pm}0.0566$ & $0.9157{\pm}0.0486$ & $0.8235{\pm}0.0214$ \\
PH2 & Generic U-Net & $0.7692{\pm}0.0467$ & $0.7174{\pm}0.0554$ & $0.5623{\pm}0.0710$ & $0.5988{\pm}0.0918$ & $0.9110{\pm}0.0300$ & $0.7061{\pm}0.0738$ & $0.7174{\pm}0.0554$ \\
PH2 & SwinUNETR & $0.8789{\pm}0.0225$ & $0.8116{\pm}0.0252$ & $0.6837{\pm}0.0361$ & $0.8181{\pm}0.0338$ & $0.8088{\pm}0.0541$ & $0.9153{\pm}0.0144$ & $0.8116{\pm}0.0252$ \\
PH2 & U-Net & $0.9313{\pm}0.0154$ & $0.8900{\pm}0.0182$ & $0.8022{\pm}0.0292$ & $0.9282{\pm}0.0246$ & $0.8570{\pm}0.0444$ & $0.9679{\pm}0.0122$ & $0.8900{\pm}0.0182$ \\
PH2 & MLFFM-SegDiff & $\mathbf{0.9591{\pm}0.0220}$ & $\mathbf{0.9384{\pm}0.0269}$ & $\mathbf{0.8852{\pm}0.0461}$ & $\mathbf{0.9424{\pm}0.0456}$ & $\mathbf{0.9354{\pm}0.0132}$ & $\mathbf{0.9697{\pm}0.0287}$ & $\mathbf{0.9384{\pm}0.0269}$ \\
\midrule
HAM10000 & DermoSegDiff & $0.9491{\pm}0.0048$ & $0.9002{\pm}0.0115$ & $0.8187{\pm}0.0187$ & $0.9408{\pm}0.0137$ & $0.8639{\pm}0.0300$ & $0.9800{\pm}0.0052$ & $0.9002{\pm}0.0115$ \\
HAM10000 & Generic U-Net & $0.8491{\pm}0.0146$ & $0.7363{\pm}0.0098$ & $0.5828{\pm}0.0122$ & $0.6933{\pm}0.0405$ & $0.7899{\pm}0.0401$ & $0.8704{\pm}0.0319$ & $0.7363{\pm}0.0098$ \\
HAM10000 & SwinUNETR & $0.9640{\pm}0.0005$ & $0.9320{\pm}0.0009$ & $0.8727{\pm}0.0016$ & $0.9379{\pm}0.0029$ & $0.9263{\pm}0.0044$ & $0.9777{\pm}0.0018$ & $0.9320{\pm}0.0009$ \\
HAM10000 & U-Net & $0.9647{\pm}0.0007$ & $0.9334{\pm}0.0014$ & $0.8752{\pm}0.0025$ & $0.9389{\pm}0.0056$ & $0.9280{\pm}0.0038$ & $0.9780{\pm}0.0018$ & $0.9334{\pm}0.0014$ \\
HAM10000 & MLFFM-SegDiff & $\mathbf{0.9671{\pm}0.0010}$ & $\mathbf{0.9361{\pm}0.0015}$ & $\mathbf{0.8800{\pm}0.0025}$ & $\mathbf{0.9437{\pm}0.0041}$ & $\mathbf{0.9286{\pm}0.0022}$ & $\mathbf{0.9806{\pm}0.0015}$ & $\mathbf{0.9361{\pm}0.0014}$ \\
\bottomrule
\end{tabular}%
}
\end{table*}

To further summarize the overall performance, the average results across the three datasets are reported in Table~\ref{tab:average-results}.
MLFFM-SegDiff achieves the highest average Accuracy, F1-score, Jaccard index, Recall, and Dice coefficient.
U-Net obtains slightly higher average Precision and Specificity, indicating more conservative foreground prediction.
In contrast, MLFFM-SegDiff shows stronger lesion coverage and overlap quality.

\begin{table}[t]
\centering
\caption{Average performance across ISIC2018, PH2, and HAM10000.}
\label{tab:average-results}
\resizebox{\linewidth}{!}{%
\begin{tabular}{lccccccc}
\toprule
Model & Acc. & F1 & Jaccard & Precision & Recall & Specificity & Dice \\
\midrule
DermoSegDiff & $0.9079$ & $0.8397$ & $0.7317$ & $0.8692$ & $0.8377$ & $0.9340$ & $0.8397$ \\
Generic U-Net & $0.8226$ & $0.7066$ & $0.5483$ & $0.6434$ & $0.8009$ & $0.8222$ & $0.7066$ \\
SwinUNETR & $0.9305$ & $0.8739$ & $0.7798$ & $0.8823$ & $0.8671$ & $0.9547$ & $0.8739$ \\
U-Net & $0.9481$ & $0.9006$ & $0.8202$ & $\mathbf{0.9171}$ & $0.8859$ & $\mathbf{0.9716}$ & $0.9006$ \\
MLFFM-SegDiff & $\mathbf{0.9589}$ & $\mathbf{0.9207}$ & $\mathbf{0.8546}$ & $0.9150$ & $\mathbf{0.9280}$ & $0.9697$ & $\mathbf{0.9207}$ \\
\bottomrule
\end{tabular}%
}
\end{table}

\subsection{Ablation Study}
Dice and sensitivity are selected as the primary metrics for ablation analysis, since the proposed modules are designed to improve lesion-region coverage and boundary-aware segmentation.
Here, D-U-Net denotes the proposed dual-path U-Net encoder, MLFFM denotes the multi-level feature fusion module, and Edge Loss denotes the boundary-sensitive loss term.

\begin{table}[htbp]
\centering
\caption{Ablation study of the proposed modules on three validation sets.}
\label{tab:ablation_all}
\small
\setlength{\tabcolsep}{4pt}
\begin{tabularx}{\linewidth}{
>{\centering\arraybackslash}p{1.7cm}
>{\centering\arraybackslash}X
>{\centering\arraybackslash}X
>{\centering\arraybackslash}X
cc}
\toprule
\textbf{Dataset} & \textbf{D-U-Net} & \textbf{MLFFM} & \textbf{Edge Loss} & \textbf{Dice} & \textbf{Sens.} \\
\midrule
\multirow{5}{*}{ISIC2018}
 &  &  &  & 0.8715 & 0.8330 \\
 & $\checkmark$ &  &  & \textbf{0.8942} & 0.8690 \\
 &  & $\checkmark$ &  & 0.8889 & 0.8767 \\
 &  &  & $\checkmark$ & 0.8931 & 0.8778 \\
 & $\checkmark$ & $\checkmark$ & $\checkmark$ & 0.8877 & \textbf{0.9199} \\
\midrule
\multirow{5}{*}{HAM10000}
 &  &  &  & 0.9177 & 0.9198 \\
 & $\checkmark$ &  &  & 0.9276 & 0.9213 \\
 &  & $\checkmark$ &  & 0.9240 & 0.9247 \\
 &  &  & $\checkmark$ & 0.9301 & 0.9203 \\
 & $\checkmark$ & $\checkmark$ & $\checkmark$ & \textbf{0.9361} & \textbf{0.9286} \\
\midrule
\multirow{5}{*}{PH2}
 &  &  &  & 0.9117 & 0.8732 \\
 & $\checkmark$ &  &  & 0.9301 & 0.8877 \\
 &  & $\checkmark$ &  & 0.9299 & 0.8790 \\
 &  &  & $\checkmark$ & 0.9275 & 0.9023 \\
 & $\checkmark$ & $\checkmark$ & $\checkmark$ & \textbf{0.9384} & \textbf{0.9354} \\
\bottomrule
\end{tabularx}
\end{table}

As shown in Table~\ref{tab:ablation_all}, the proposed components generally improve lesion coverage and boundary sensitivity.
On HAM10000 and PH2, the full configuration achieves the best Dice and sensitivity, indicating that D-U-Net, MLFFM, and Edge Loss are complementary.
On ISIC2018, the full model obtains the highest sensitivity, while the D-U-Net-only variant achieves the highest Dice score.
This suggests that the complete configuration tends to cover lesion regions more aggressively, whereas the D-U-Net-only variant provides a more balanced prediction in this validation setting.
Overall, the ablation results support the effectiveness of the proposed design, particularly from the perspective of recall-oriented lesion coverage.

\subsection{Visual Comparison}

\begin{figure*}[htb]
  \centering
  \includegraphics[width=\linewidth]{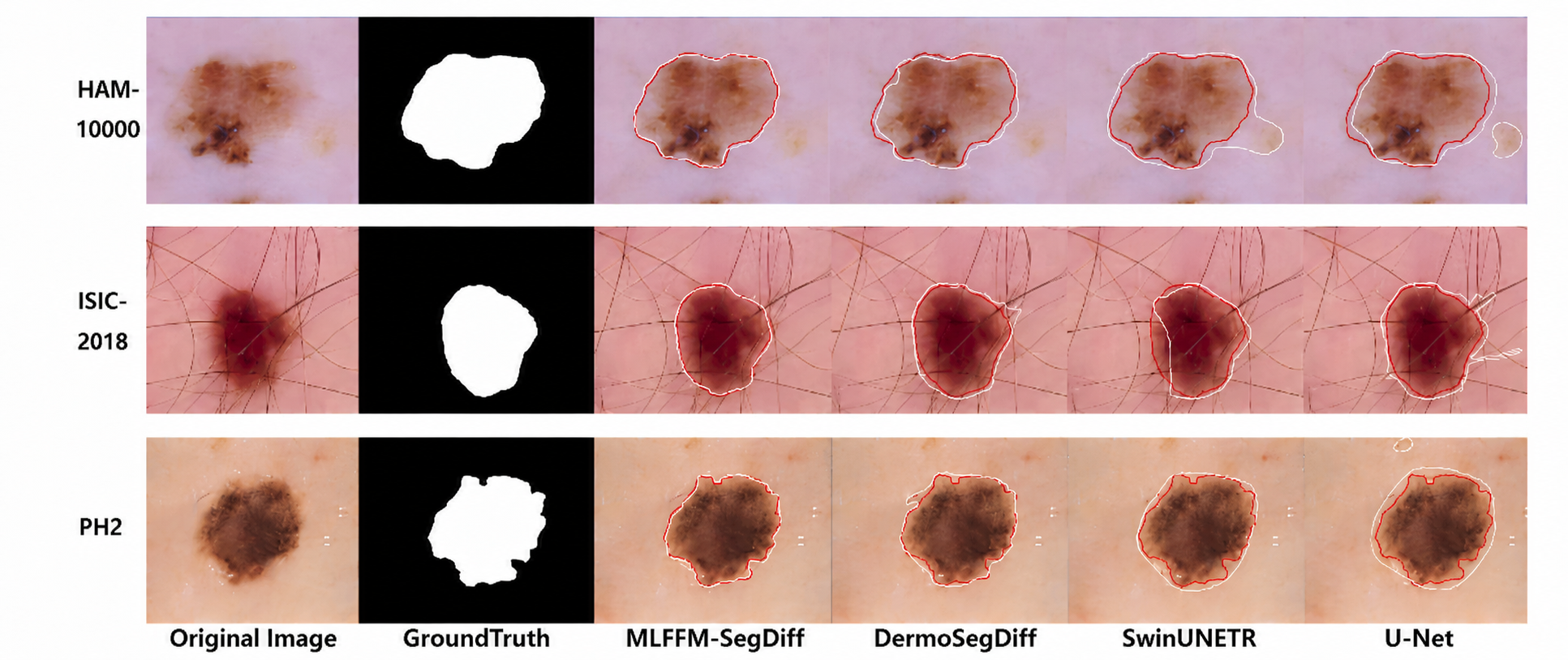}
    \caption{
        Visual comparison of segmentation results on three datasets. The white contours represent model predictions, while the red contours represent ground-truth lesion boundaries.
    }
  \label{fig:display}
\end{figure*}

Representative segmentation results are shown in Fig.~\ref{fig:display}.
The white contours represent model predictions, while the red contours represent ground-truth lesion boundaries.
Compared with the baseline methods, MLFFM-SegDiff generally produces lesion masks that better follow the target regions, especially in cases with irregular boundaries or ambiguous lesion appearance.

\subsection{Result Analysis}

On ISIC2018, MLFFM-SegDiff obtains the best overlap-related results, including F1-score, Jaccard index, Recall, and Dice coefficient.
The higher Recall indicates that the proposed method can cover lesion regions more completely.
However, its Precision and Specificity are slightly lower than the best comparison results, suggesting that improved lesion coverage may introduce some additional foreground predictions.

On PH2, MLFFM-SegDiff achieves the best results on all evaluation metrics.
Compared with U-Net, the Jaccard score improves from 0.8022 to 0.8852, and the Dice score improves from 0.8900 to 0.9384.
This demonstrates that the proposed method is effective on relatively small-scale dermoscopic datasets, where boundary modeling and feature fusion are particularly important.

On HAM10000, the baseline methods already achieve strong segmentation performance.
Nevertheless, MLFFM-SegDiff still obtains consistent improvements across all metrics, showing that the proposed design remains effective on larger-scale dermoscopic image data.

Overall, the main advantage of MLFFM-SegDiff lies in lesion-region coverage and overlap quality.
By enhancing the interaction between noisy mask features and image guidance features, and by fusing multi-level skip features before decoding, the proposed method better combines shallow boundary details with deep semantic information, thereby improving lesion mask reconstruction.

\section{Conclusion}

In this paper, we proposed MLFFM-SegDiff, a diffusion-based model for skin lesion segmentation.
The proposed method improves the denoising network through dual-path feature interaction, multi-level skip feature fusion, and configurable boundary-sensitive optimization.
The dual-path encoder enhances the interaction between noisy mask features and dermoscopic image guidance features, while the Multi-Level Feature Fusion Module (MLFFM) strengthens skip connections by combining channel attention, spatial attention, scale alignment, and cross-level fusion.
With these designs, the decoder can better integrate shallow boundary details and deep semantic representations for lesion mask reconstruction.

Experiments on ISIC2018, PH2, and HAM10000 demonstrate the effectiveness of the proposed method.
Compared with DermoSegDiff, Generic U-Net, SwinUNETR, and U-Net, MLFFM-SegDiff achieves strong overall segmentation performance and obtains the best average Accuracy, F1-score, Jaccard index, Recall, and Dice coefficient across the three datasets.
In particular, the average Jaccard index and Dice coefficient reach 0.8546 and 0.9207, respectively, showing improved lesion-region coverage and overlap quality.

Although promising results are achieved, several limitations remain.
On ISIC2018, Precision and Specificity are not the highest among the compared methods, indicating that background suppression and fine-grained boundary control can be further improved.
In addition, diffusion-based segmentation requires iterative denoising during inference, which may lead to higher computational cost than conventional feed-forward segmentation networks.
Future work will focus on efficient diffusion sampling, stronger boundary and background constraints, and broader validation on larger-scale, multi-center, and high-resolution clinical datasets.
Extending the framework to other medical image segmentation tasks will also be investigated to further evaluate its generalization ability and clinical application potential.

\section*{Acknowledgements}

This work was supported by the Science Foundation of Keyi College of Zhejiang Sci-Tech University under Grant No. KYQD251104 and  the  Noncommunicable Chronic Diseases-National Science and Technology Major Project under Grant No. 2025ZD0544500. 

\newpage

\bibliographystyle{elsarticle-num}
\bibliography{references}

@inproceedings{long2015fcn,
  title={Fully Convolutional Networks for Semantic Segmentation},
  author={Long, Jonathan and Shelhamer, Evan and Darrell, Trevor},
  booktitle={Proceedings of the IEEE Conference on Computer Vision and Pattern Recognition (CVPR)},
  pages={3431--3440},
  year={2015},
  doi={10.1109/CVPR.2015.7298965}
}

@inproceedings{ronneberger2015unet,
  title={U-Net: Convolutional Networks for Biomedical Image Segmentation},
  author={Ronneberger, Olaf and Fischer, Philipp and Brox, Thomas},
  booktitle={Medical Image Computing and Computer-Assisted Intervention (MICCAI)},
  pages={234--241},
  year={2015},
  doi={10.1007/978-3-319-24574-4_28}
}

@article{oktay2018attunet,
  title={Attention U-Net: Learning Where to Look for the Pancreas},
  author={Oktay, Ozan and Schlemper, Jo and others},
  journal={arXiv preprint arXiv:1804.03999},
  year={2018}
}

@inproceedings{zhou2018unetpp,
  title={UNet++: A Nested U-Net Architecture for Medical Image Segmentation},
  author={Zhou, Zongwei and Siddiquee, Md Mahfuzur Rahman and Tajbakhsh, Nima and Liang, Jianming},
  booktitle={Deep Learning in Medical Image Analysis and Multimodal Learning for Clinical Decision Support (DLMIA)},
  pages={3--11},
  year={2018},
  doi={10.1007/978-3-030-00889-5_1}
}

@article{isensee2021nnunet,
  title={nnU-Net: A Self-configuring Method for Deep Learning-based Biomedical Image Segmentation},
  author={Isensee, Fabian and Jaeger, Paul F. and Kohl, Simon A. A. and Petersen, Jens and Maier-Hein, Klaus H.},
  journal={Nature Methods},
  volume={18},
  pages={203--211},
  year={2021},
  doi={10.1038/s41592-020-01008-z}
}

@article{chen2021transunet,
  title={TransUNet: Transformers Make Strong Encoders for Medical Image Segmentation},
  author={Chen, Jieneng and Lu, Yongyi and Yu, Qihang and others},
  journal={arXiv preprint arXiv:2102.04306},
  year={2021}
}

@article{hatamizadeh2022swinunetr,
  title={Swin UNETR: Swin Transformers for Semantic Segmentation of Brain Tumors in MRI Images},
  author={Hatamizadeh, Ali and Xu, Dong and Myronenko, Andriy and others},
  journal={arXiv preprint arXiv:2201.01266},
  year={2022}
}

@inproceedings{ho2020ddpm,
  title={Denoising Diffusion Probabilistic Models},
  author={Ho, Jonathan and Jain, Ajay and Abbeel, Pieter},
  booktitle={Advances in Neural Information Processing Systems (NeurIPS)},
  pages={6840--6851},
  year={2020}
}

@inproceedings{song2021ddim,
  title={Denoising Diffusion Implicit Models},
  author={Song, Jiaming and Meng, Chenlin and Ermon, Stefano},
  booktitle={International Conference on Learning Representations (ICLR)},
  year={2021}
}

@inproceedings{nichol2021improved,
  title={Improved Denoising Diffusion Probabilistic Models},
  author={Nichol, Alex and Dhariwal, Prafulla},
  booktitle={International Conference on Machine Learning (ICML)},
  pages={8162--8171},
  year={2021}
}

@article{segdiff2021,
  title={SegDiff: Image Segmentation with Diffusion Probabilistic Models},
  author={Amit, Tomer and others},
  journal={arXiv preprint arXiv:2112.00390},
  year={2021}
}

@article{medsegdiff2022,
  title={MedSegDiff: Medical Image Segmentation with Diffusion Probabilistic Model},
  author={Wu, Junde and others},
  journal={arXiv preprint arXiv:2211.00611},
  year={2022}
}

@article{dermosegdiff2023,
  title={DermoSegDiff: A Boundary-Aware Segmentation Diffusion Model for Skin Lesion Delineation},
  author={Bozorgpour, Afshin and others},
  journal={arXiv preprint arXiv:2308.02959},
  year={2023}
}

@inproceedings{lin2017fpn,
  title={Feature Pyramid Networks for Object Detection},
  author={Lin, Tsung-Yi and Dollar, Piotr and Girshick, Ross and He, Kaiming and Hariharan, Bharath and Belongie, Serge},
  booktitle={Proceedings of the IEEE Conference on Computer Vision and Pattern Recognition (CVPR)},
  pages={2117--2125},
  year={2017},
  doi={10.1109/CVPR.2017.106}
}

@inproceedings{woo2018cbam,
  title={CBAM: Convolutional Block Attention Module},
  author={Woo, Sanghyun and Park, Jongchan and Lee, Joon-Young and Kweon, In So},
  booktitle={European Conference on Computer Vision (ECCV)},
  pages={3--19},
  year={2018}
}

@article{kervadec2021boundary,
  title={Boundary Loss for Highly Unbalanced Segmentation},
  author={Kervadec, Hoel and Bouchtala, Sahar and others},
  journal={Medical Image Analysis},
  volume={67},
  pages={101851},
  year={2021},
  doi={10.1016/j.media.2020.101851}
}

@article{codella2019isic,
  title={Skin Lesion Analysis Toward Melanoma Detection 2018: A Challenge Dataset},
  author={Codella, Noel C. F. and Gutman, David and Celebi, Emre and others},
  journal={arXiv preprint arXiv:1902.03368},
  year={2019}
}

@article{tschandl2018ham10000,
  title={The HAM10000 Dataset: A Large Collection of Multi-Source Dermatoscopic Images of Common Pigmented Skin Lesions},
  author={Tschandl, Philipp and Rosendahl, Cliff and Kittler, Harald},
  journal={Scientific Data},
  volume={5},
  pages={180161},
  year={2018},
  doi={10.1038/sdata.2018.161}
}

@inproceedings{mendonca2013ph2,
  title={PH2: A Dermoscopic Image Database for Research and Benchmarking},
  author={Mendonça, Teresa and Ferreira, Pedro M. and Marques, Jorge S. and others},
  booktitle={IEEE International Conference on Engineering in Medicine and Biology Society (EMBC)},
  pages={5437--5440},
  year={2013},
  doi={10.1109/EMBC.2013.6610779}
}

@article{brinker2019,
  title={Deep Learning Outperformed Dermatologists in Melanoma Classification},
  author={Brinker, Titus J. and Hekler, Achim and others},
  journal={European Journal of Cancer},
  volume={119},
  pages={93--100},
  year={2019}
}

@article{esteva2017,
  title={Dermatologist-level classification of skin cancer with deep neural networks},
  author={Esteva, Andre and Kuprel, Brett and Novoa, Roberto and others},
  journal={Nature},
  volume={542},
  pages={115--118},
  year={2017},
  doi={10.1038/nature21056}
}

@article{celebi2019,
  title={Dermoscopic Image Analysis: Overview and Future Directions},
  author={Celebi, Emre and Wen, Qian},
  journal={IEEE Reviews in Biomedical Engineering},
  year={2019}
}

@article{dosovitskiy2021,
  title={An Image is Worth 16x16 Words: Transformers for Image Recognition at Scale},
  author={Dosovitskiy, Alexey and Beyer, Lucas and Kolesnikov, Alexander and others},
  journal={arXiv preprint arXiv:2010.11929},
  year={2021}
}

\end{document}